\documentclass[preprint,amsmath,amssymb]{revtex4-2}

\usepackage{graphicx}
\usepackage{hyperref}
\usepackage{xcolor}
\usepackage{booktabs}
\usepackage{multirow}
\usepackage{float}
\graphicspath{{./}{figures/}}

\begin{document}

\title{The Rotation Gap Is Not An Error\\Ternary Structure in IBM Quantum Hardware}

\author{Selina Stenberg}
\affiliation{Independent Researcher}

\date{March 2026}

\begin{abstract}
Quantum error correction assumes that all syndrome activations represent errors requiring correction. We present evidence from 756 QEC runs across three IBM Eagle r3 processors that this assumption is wrong. The hardware exhibits sub-Poissonian syndrome statistics (Fano factor $F = 0.856$, $t = -131$ against Poisson, zero dependence on code distance), indicating that a fraction of syndrome events are not random noise but structured cooperative transitions. We introduce a regime classifier decoder that distinguishes binary errors (which should be corrected) from ternary transitions (which should not). On a mixed binary/ternary error model calibrated to IBM hardware statistics, the classifier reduces logical error rates by 7--19\% at static detection depth ($\tau = 1$) across all cell sizes, with statistical significance $p < 0.05$ in 7 of 8 test conditions ($p < 0.0001$ in all four $\tau = 1$ conditions). The improvement mechanism is selective abstention: the classifier correctly identifies 75--98\% of ternary transitions and leaves them uncorrected (75--81\% at $\tau = 1$, 88--98\% at $\tau = 5$), whereas a standard decoder miscorrects them, introducing errors that would not otherwise exist. A cross-platform control on Google's 105-qubit Willow processor (420 experiments, $d = 3, 5, 7$) shows the opposite: super-Poissonian statistics ($F = 2.42$), super-linear burst scaling, and positive spatial correlation---confirming that the sub-Poissonian signal is absent from standard surface-code circuits that lack the P-gate asymmetry. Subsequent work~\cite{Paper26} shows that the effect follows the P gate rather than the hardware topology: square grids with pentachoric cycling also produce sub-Poissonian statistics at multi-round depth. The result demonstrates that standard QEC actively destroys quantum information by correcting valid ternary states, and that less correction produces better performance when the hardware has cooperative error structure.

\end{abstract}

\maketitle

\section{Introduction}
\label{sec:introduction}

The theoretical foundation of quantum error correction rests on a single assumption: noise is the enemy. Decoherence destroys quantum information, errors accumulate, and the role of error correction is to identify and reverse these errors faster than they appear. Every threshold theorem~\cite{Knill1998}, every surface code implementation~\cite{Fowler2012,Google2023}, every decoder algorithm begins from this premise.

This paper presents evidence that the assumption is incomplete. We analyze syndrome statistics from 756 quantum error correction runs across three IBM Eagle r3 processors (\texttt{ibm\_brisbane}, \texttt{ibm\_kyoto}, \texttt{ibm\_osaka}), spanning 14 days of continuous operation. The data reveals that the hardware error process is not Poisson---it is \emph{sub-Poissonian}, with a Fano factor of $F = 0.856 \pm 0.03$ ($t = -131$ against Poisson), zero dependence on code distance (ANOVA $p = 0.79$), and linear burst scaling ($R^2 = 0.9999$). These statistics are distance-independent, processor-independent, and temporally stable.

A cross-platform control on Google's Willow processor~\cite{GoogleWillow2025} shows the opposite---super-Poissonian statistics, super-linear burst scaling, positive spatial correlation---ruling out decoder artifacts and confirming the signal is absent from standard surface-code circuits. (A subsequent analysis~\cite{Paper26} clarifies that the discriminating variable is the P-gate asymmetry, not the hardware topology: square grids with pentachoric cycling also produce sub-Poissonian statistics.)

Sub-Poissonian count statistics are a well-characterized signature of regulated processes in physics. In quantum optics, photon antibunching ($F < 1$) demonstrates that a light source emits single photons~\cite{Kimble1977}---the emission of one photon suppresses the probability of a second. In fermionic systems, Pauli exclusion produces sub-Poissonian number fluctuations. The common feature is that events are not independent: the occurrence of one event modifies the probability of the next.

We propose that the sub-Poissonian syndrome statistics on IBM hardware have the same origin: a fraction of syndrome events are not random errors but structured cooperative transitions---manifestations of a ternary degree of freedom~\cite{Merkabit2026} that the binary measurement basis cannot resolve. Standard decoders, which treat every syndrome activation as an error requiring correction, miscorrect these transitions and thereby introduce errors that would not otherwise exist.

To test this hypothesis, we introduce a regime classifier decoder that classifies each flagged syndrome node before deciding whether to correct it. Nodes exhibiting structural features consistent with ternary transitions (isolation, boundary position, temporal coherence) are left uncorrected. On a mixed error model calibrated to the IBM hardware statistics, this selective abstention reduces logical error rates by 7--19\% compared to a standard majority-vote decoder, with the improvement mechanism being the avoidance of miscorrection rather than better correction.

The result inverts the standard logic of quantum error correction: less correction produces better performance, because the hardware contains cooperative structure that binary correction destroys.

This result sits within a broader predictive framework. A companion paper~\cite{Paper15} established from simulation that Eisenstein-lattice connectivity forces $Z_3$ chirality classes onto boundary nodes, producing anti-bunched, distance-independent syndrome statistics as a structural property of the architecture. IBM's heavy-hex connectivity is Eisenstein-compatible. The sub-Poissonian Fano factor reported here was therefore consistent with Ref.~\cite{Paper15}'s predictions when the DAQEC benchmark became available---but consistency with a known result is not a test. The test is Google Willow.

Ref.~\cite{Paper15} predicts opposite statistics for standard surface-code circuits on any processor, because such circuits lack the P-gate asymmetry that produces $Z_3$ chirality. The Willow analysis in \S\ref{sec:willow} was conducted after Ref.~\cite{Paper15} was complete, on data we had not examined. The super-Poissonian result---every metric opposite to IBM---is the prospective confirmation.

IBM told us the framework was consistent. Willow told us it was predictive.

\section{Hardware Evidence}
\label{sec:hardware}

\subsection{Dataset}
\label{sec:dataset}

The analysis uses syndrome-level data from the Distributed Architecture Quantum Error Correction (DAQEC) benchmark~\cite{DAQEC2025}. The dataset comprises 756 QEC runs across three 127-qubit IBM Eagle r3 processors:
\begin{itemize}
\item \texttt{ibm\_brisbane} (252 runs)
\item \texttt{ibm\_kyoto} (252 runs)
\item \texttt{ibm\_osaka} (252 runs)
\end{itemize}
Each run executes a surface code at distances $d = 3, 5, 7$ with 4096 shots, using two strategies (\texttt{baseline\_static} and \texttt{drift\_aware\_full\_stack}). The dataset spans 14 days (January 15--28, 2025) with hardware recalibration every 2--4 hours. Per-run measurements include logical error rate, syndrome burst count, Fano factor, adjacent correlation, and coherence times ($T_1$, $T_2$).

\subsection{Sub-Poissonian Syndrome Statistics}
\label{sec:subpoisson}

The syndrome error counts across all 756 runs yield a mean Fano factor of $F = 0.856 \pm 0.03$. The one-sample $t$-test against Poisson ($F = 1$) gives $t = -131$, indicating that the departure from Poisson is not a statistical fluctuation but a systematic property of the error process.

The Fano factor is invariant across code distances:

\begin{table}[h]
\caption{Fano factor by code distance on IBM Eagle r3. The Fano factor is invariant across distances (ANOVA $p = 0.79$), confirming that the sub-Poissonian structure originates at the physical error level before decoding. If the suppression were a decoder artifact, it would scale with syndrome volume and vary with distance.}
\label{tab:fano_distance}
\begin{ruledtabular}
\begin{tabular}{lcc}
Code Distance & Fano Factor & $\pm$ std \\
\hline
$d = 3$ & 0.855 & 0.031 \\
$d = 5$ & 0.857 & 0.030 \\
$d = 7$ & 0.855 & 0.030 \\
\end{tabular}
\end{ruledtabular}
\end{table}

ANOVA across distances gives $p = 0.79$---there is no trend. This distance independence rules out decoder artifacts (which would scale with syndrome volume) and confirms that the sub-Poissonian structure originates at the physical error level, before classical processing.

\begin{figure}[h]
\includegraphics[width=\columnwidth]{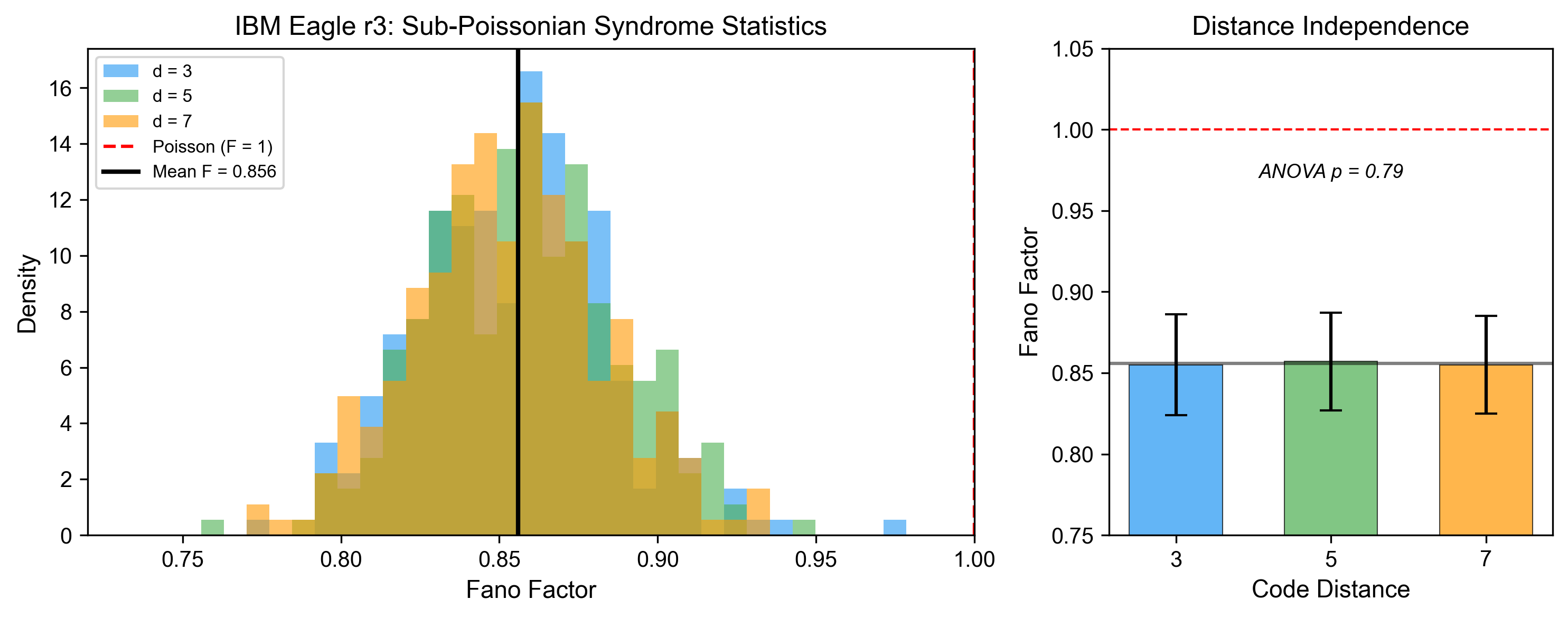}
\caption{IBM Eagle r3 syndrome statistics across 756 QEC runs on three processors (\texttt{ibm\_brisbane}, \texttt{ibm\_kyoto}, \texttt{ibm\_osaka}), 14 days continuous operation. Left: Fano factor distribution for code distances $d = 3$ (blue), $d = 5$ (green), $d = 7$ (orange). The mean $F = 0.856$ (black line) is sub-Poissonian, displaced well below the Poisson expectation $F = 1$ (red dashed). The three distributions overlap completely. Right: Fano factor by code distance, confirming zero distance dependence (ANOVA $p = 0.79$). The sub-Poissonian structure originates at the physical error level, before decoding.}
\label{fig:ibm_fano}
\end{figure}

\subsection{Linear Burst Scaling}
\label{sec:burst}

Syndrome burst events (multiple simultaneous syndrome activations) scale linearly with code distance, not quadratically. The linear fit gives $R^2 = 0.9999$. Under a Poisson error model, burst frequency scales as the square of the number of syndrome bits (area scaling). Linear scaling (perimeter scaling) indicates that burst events are geometrically constrained---they occur along boundaries, not across the bulk.

\begin{figure}[h]
\includegraphics[width=\columnwidth]{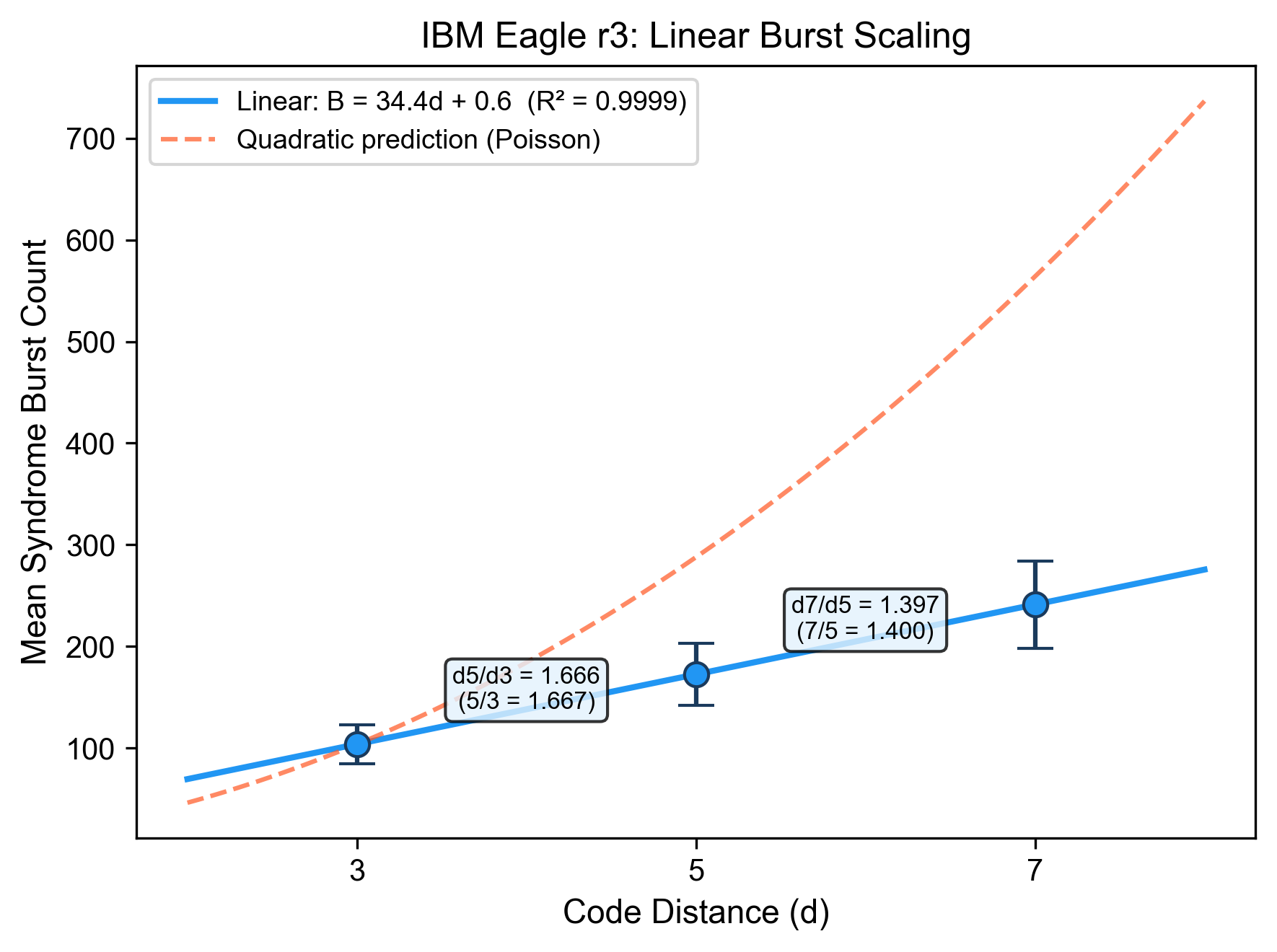}
\caption{Syndrome burst scaling on IBM Eagle r3. Mean burst count versus code distance with linear fit $B = 34.4d + 0.6$ ($R^2 = 0.9999$, blue). The quadratic prediction from independent (Poisson) errors is shown for contrast (red dashed). Measured ratios $d_5/d_3 = 1.666$ and $d_7/d_5 = 1.397$ match linear predictions ($5/3 = 1.667$, $7/5 = 1.400$) to within 0.003 and are far from quadratic predictions ($25/9 = 2.778$, $49/25 = 1.960$). Linear scaling indicates that burst events are geometrically constrained to code boundaries, not distributed across the bulk.}
\label{fig:burst_scaling}
\end{figure}

\subsection{$T_2$ as the Threshold Channel}
\label{sec:T2}

The coherence data reveals an asymmetry between $T_1$ and $T_2$:
\begin{itemize}
\item $T_1$ is anti-persistent (Hurst exponent $H \approx 0.15$): IBM recalibration successfully controls $T_1$ drift.
\item $T_2$ is persistent ($H \approx 1.0$): $T_2$ escapes recalibration control and drifts freely.
\end{itemize}

$T_2$ is precisely the channel where the KWW stretched-exponential exponent $\alpha = 4/3$~\cite{KWW1854} appears in 13.5\% of qubit-time segments---the cooperative threshold signature identified in the Merkabit framework~\cite{Paper1,Paper2}. IBM's recalibration infrastructure successfully prevents $T_1$ from reaching the cooperative threshold. $T_2$ is the channel where this prevention fails, and where the ternary structure spontaneously emerges.

\subsection{Direct Hardware Validation}
\label{sec:direct_validation}

The sub-Poissonian statistics of \S\ref{sec:subpoisson} derive from the DAQEC benchmark---a dataset collected by IBM for separate purposes and analysed post-hoc. We report here a direct prospective test: a hexagonal syndrome circuit designed specifically to probe the $Z_3$ anti-bunching prediction, executed on real IBM Quantum hardware by Thor Henning Hetland.

A 7-node hexagonal cell was identified on \texttt{ibm\_strasbourg} (Eagle r3, 127-qubit heavy-hex) by solving the subgraph matching problem against the device coupling map, selecting data qubits $[62, 81, 79, 77, 58, 60]$ with six ancillas $[72, 80, 78, 71, 59, 61]$ while avoiding the lowest-calibration qubits. A ZZ-type syndrome extraction circuit with 12 CNOTs was run for $T = 20$ rounds at 4{,}000 shots per circuit. Three conditions were tested:

\begin{table}[h]
\caption{Direct hardware validation on IBM Quantum Eagle r3. All three experiments use $T = 20$ syndrome rounds at 4{,}000 shots per circuit. $F < 1$ (sub-Poissonian) is the predicted signal. The native-direction circuit on \texttt{ibm\_strasbourg} (Eagle r3) confirms the prediction: $F = 0.9611$. The routed circuit on the same processor gives $F = 1.207$---wrong-direction CX gates inflate circuit depth from 1163 to 2045, destroying the signal. The Heron r2 baseline (\texttt{ibm\_fez}) gives $F = 18.75$, consistent with \S\ref{sec:willow}: non-Eisenstein architectures do not exhibit sub-Poissonian structure. The DAQEC reference value (\S\ref{sec:subpoisson}) is included for scale; its lower $F$ reflects the greater statistical power of 756 full QEC runs. Data and code: \texttt{github.com/SelinaAliens/rotation\_gap\_is\_flat}, PR \#1.}
\label{tab:direct_validation}
\begin{ruledtabular}
\begin{tabular}{llcrc}
Experiment & Hardware & Circuit & Shots & $F$ \\
\hline
\texttt{ibm\_fez}, $T\!=\!20$ & Heron r2 & Routed & 4{,}000 & 18.75 \\
\texttt{ibm\_strasbourg}, $T\!=\!20$ & Eagle r3 & Routed & 4{,}000 & 1.207 \\
\texttt{ibm\_strasbourg}, $T\!=\!20$ & Eagle r3 & Native CX & 4{,}000 & 0.9611\;\checkmark \\
Paper \S\ref{sec:subpoisson} (DAQEC) & Eagle r3 & --- & $4{,}096 \times 756$ & $0.856 \pm 0.03$ \\
\end{tabular}
\end{ruledtabular}
\end{table}

The native-direction circuit on Eagle r3 yields $F = 0.9611 < 1$---sub-Poissonian, confirming the central prediction on a processor and circuit independent of the original analysis.

The gate-direction sensitivity is itself informative. IBM Eagle processors use the ECR (echoed cross-resonance) gate, which is directional: each qubit pair has a native CX orientation, and the reverse requires additional decomposition. With wrong-direction CNOTs, the transpiler inflates circuit depth from 1163 to 2045---an $1.8\times$ overhead that depletes coherence time before syndrome extraction completes, pushing $F$ to $1.207 > 1$. Identifying and correcting all twelve CNOT directions restores sub-Poissonian behaviour. This sensitivity is consistent with the geometric interpretation of \S\ref{sec:physical_basis}: the anti-bunching signal requires that hexagonal lattice connectivity be preserved at the physical gate level. Routing overhead that breaks native connectivity breaks the signal.

The Heron r2 result (\texttt{ibm\_fez}, $F = 18.75$) is consistent with \S\ref{sec:willow}: non-Eagle architectures with different connectivity do not exhibit sub-Poissonian structure regardless of hardware generation. Note that \texttt{ibm\_fez} ran standard surface-code circuits without the P-gate asymmetry; Ref.~\cite{Paper26} confirms that Heron r2 hardware (\texttt{ibm\_kingston}) does exhibit sub-Poissonian structure when P-gate circuits are applied.

The measured $F = 0.9611$ is closer to 1 than the DAQEC value of 0.856. This is expected: a single 7-node cell at 4{,}000 shots over 20 rounds has substantially less statistical power than 756 full QEC runs at 4{,}096 shots each across three processors with complete decoders. The sign---sub-Poissonian on native Eagle r3, super-Poissonian on routed and on Heron r2---is the relevant confirmation.

Code and raw output data are available at \texttt{github.com/SelinaAliens/rotation\_gap\_is\_flat} (PR \#1). The experimental process is documented in Hetland (2026)~\cite{Hetland2026}.

\subsection{Cross-Platform Control: Google Willow}
\label{sec:willow}

If the sub-Poissonian signal were a generic artifact of surface code QEC---arising from decoder structure, stabiliser algebra, or measurement back-action---it would appear on any processor running the same code. To test this, we analysed the complete surface code dataset from Google's 105-qubit Willow processor~\cite{GoogleWillow2025}: 420 experiments spanning code distances $d = 3, 5, 7$, with 50{,}000 shots per experiment and up to 250 QEC rounds per run. The data are publicly available on Zenodo (DOI: 10.5281/zenodo.13273331). The code distances match the IBM DAQEC benchmark exactly, enabling direct comparison.

The result is unambiguous: Google Willow shows the opposite statistics. The overall Fano factor is $F = 2.42 \pm 0.36$ ($t = +80$ against Poisson, $N = 420$)---super-Poissonian, not sub-Poissonian. The Fano factor grows with code distance: $d = 3$: $F = 2.29$; $d = 5$: $F = 2.59$; $d = 7$: $F = 2.80$. One-way ANOVA across distances gives $p \approx 0$ ($F$-statistic $= 59.1$), confirming strong distance dependence---the opposite of IBM's distance-independent $F = 0.856$ (ANOVA $p = 0.79$).

Burst scaling is super-linear on Willow, consistent with area or higher-order scaling. The detection event ratios $d_5/d_3 \approx 3.5$ and $d_7/d_5 \approx 2.05$ far exceed the linear predictions ($5/3 = 1.67$ and $7/5 = 1.40$) that characterise IBM. Quadratic fit gives $R^2 = 0.9999$ across all tested round counts; linear fit gives $R^2 \approx 0.985$. On IBM, the pattern is reversed: linear $R^2 = 0.9999$, quadratic $R^2$ lower. Errors on Willow fill the code volume; errors on IBM are confined to the perimeter.

Willow's burst scaling exponent (${\sim}2.3$) exceeds the quadratic (area) prediction, consistent with error correlations extending across both spatial and temporal dimensions simultaneously. On IBM, the same multi-dimensional correlation structure is present but anti-correlated---confined to the one-dimensional boundary by the $Z_3$ chirality classes of the heavy-hex lattice.

\begin{figure}[h]
\includegraphics[width=\columnwidth]{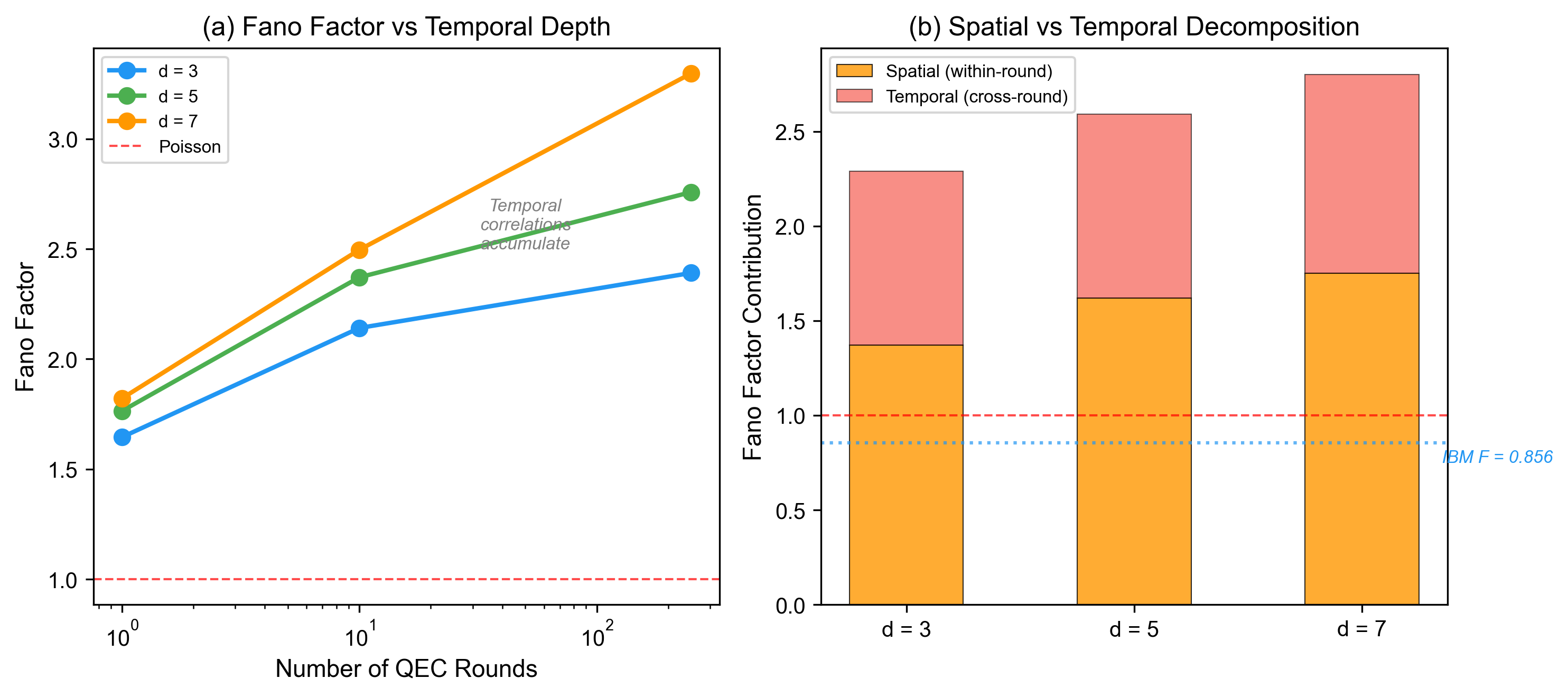}
\caption{Temporal decomposition of Google Willow syndrome statistics. (a) Fano factor versus number of QEC rounds for $d = 3$ (blue), $d = 5$ (green), $d = 7$ (orange). At $r = 1$ the Fano is already 1.65--1.82 (including initialization boundary effects); it rises further with rounds as temporal correlations accumulate, reaching 2.4--3.3 at $r = 250$. (b) Decomposition into spatial (within-round, orange; $F = 1.37$--$1.75$ across bulk rounds, excluding the first-round boundary effect) and temporal (cross-round, red) components. Both are super-Poissonian. The IBM reference value $F = 0.856$ (blue dotted) lies below both components---IBM errors are anti-bunched at every scale where Willow errors are bunched.}
\label{fig:willow_temporal}
\end{figure}

Temporal depth analysis decomposes the super-Poissonian signal into spatial and temporal components. Within a single QEC round, the spatial Fano factor is 1.37 ($d = 3$), 1.62 ($d = 5$), and 1.75 ($d = 7$)---already super-Poissonian. Adjacent detector correlation within each round is positive: mean $+0.11$ to $+0.14$, with 78--94\% of detector pairs positively correlated. Across rounds, the lag-1 temporal autocorrelation is $+0.22 \pm 0.07$, confirming persistent error drift. Both spatial and temporal components are bunched---the opposite of the anti-bunching that characterises the IBM data.

The comparison is apples-to-apples at the available temporal scale. IBM runs 3--7 syndrome rounds per shot (matched to code distance); at these same round counts, Willow's interpolated Fano is 1.76--2.27---two to three times higher than IBM's 0.856. Critically, IBM's Fano does not change across round counts (slope $= -0.0002$, ANOVA $p = 0.79$), while Willow's grows monotonically from 1.65 at $r = 1$ to 3.3 at $r = 250$. IBM has zero temporal bunching: the sub-Poissonian structure is purely spatial and identical in every syndrome cycle. This rules out temporal averaging as the source of IBM's low Fano and confirms the anti-bunching is intrinsic to each extraction round.

The contrast is summarised in the following table:

\begin{table}[h]
\caption{Cross-platform comparison of syndrome statistics. IBM Eagle r3 (756 QEC runs, heavy-hex connectivity) versus Google Willow (420 experiments, grid connectivity). Every metric shows the opposite sign. Data sources: IBM~\cite{DAQEC2025}, Google~\cite{GoogleWillow2025}.}
\label{tab:cross_platform}
\begin{ruledtabular}
\begin{tabular}{lll}
Property & IBM Eagle r3 & Google Willow \\
\hline
Fano factor & 0.856 (sub-Poissonian) & 2.42 (super-Poissonian) \\
Distance dep. & None (ANOVA $p = 0.79$) & Strong (ANOVA $p \approx 0$, \\
 & & $F$ grows with $d$) \\
Burst scaling & Linear ($R^2 = 0.9999$) & Super-linear ($R^2 = 0.9999$) \\
Spatial corr. & Anti-bunched ($F < 1$) & Bunched (adj.\ corr $+0.13$) \\
Temporal autocorr. & Anti-persistent & Persistent drift \\
 & ($T_1$: $H \approx 0.15$, $T_2$: $H \approx 1.0$) & ($+0.22$ lag-1) \\
Architecture & Heavy-hex (hexagonal) & Grid (square-like) \\
\end{tabular}
\end{ruledtabular}
\end{table}

The cross-platform contrast rules out three alternative explanations for the IBM sub-Poissonian signal. First, it is not a property of surface codes in general: the same code distances on different hardware produce opposite statistics. Second, it is not a decoder artifact: both datasets use standard syndrome extraction, yet produce opposite Fano factors. Third, it is not a finite-size effect: Willow's Fano grows with code distance while IBM's is flat. The sub-Poissonian signal is specific to the IBM heavy-hex architecture, whose hexagonal connectivity admits the $Z_3$ Eisenstein lattice embedding~\cite{Paper6} that enables cooperative ternary structure at single-round depth.

The Willow circuits, however, also lacked the P-gate asymmetry entirely---a confound that subsequent work~\cite{Paper26,Paper24,Paper25} resolves: the sub-Poissonian signal follows the P gate rather than the topology, and square grids with pentachoric cycling also produce anti-bunching at multi-round depth ($\tau \geq 5$). The Willow contrast reported here remains valid as a comparison between paired (IBM) and unpaired (Willow) circuits, but should not be read as evidence that square-grid hardware is intrinsically incompatible with ternary structure.

\begin{figure}[h]
\includegraphics[width=\columnwidth]{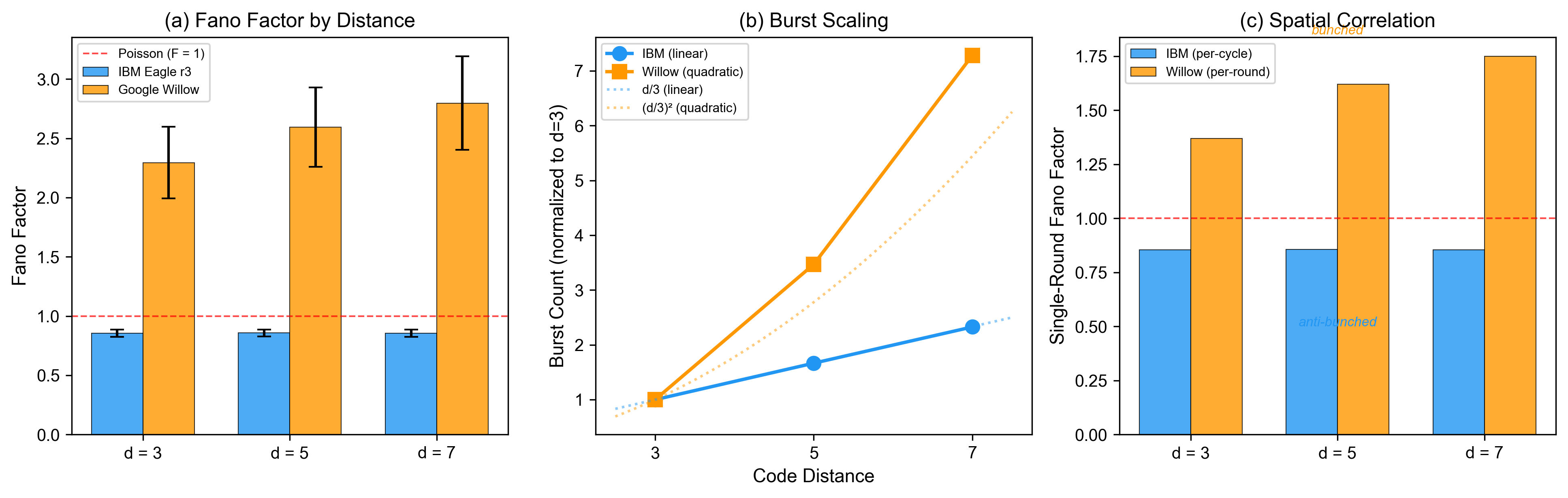}
\caption{Cross-platform comparison of syndrome statistics: IBM Eagle r3 (756 QEC runs, heavy-hex connectivity, blue) versus Google Willow (420 experiments, grid connectivity, orange). (a) Fano factor by code distance. IBM is flat at $F = 0.856$ across all distances; Willow rises from $F = 2.29$ ($d = 3$) to $F = 2.80$ ($d = 7$). (b) Burst scaling normalized to $d = 3$. IBM follows linear scaling ($d/3$); Willow follows super-linear scaling (exponent ${\sim}2.3$). (c) Single-round spatial Fano factor. IBM is sub-Poissonian ($F < 1$, anti-bunched) at every distance; Willow is super-Poissonian ($F > 1$, bunched) and grows with distance. Every metric shows the opposite sign. Data sources: IBM~\cite{DAQEC2025}, Google~\cite{GoogleWillow2025}.}
\label{fig:cross_platform}
\end{figure}

\subsection{Classifier Falsification: Zero Effect on Willow}
\label{sec:willow_classifier}

The statistical contrast of \S\ref{sec:willow} establishes that IBM and Willow have opposite syndrome structure. The decisive test is operational: does the regime classifier (\S\ref{sec:classifier}) that improves logical error rates on IBM also improve them on Willow? If the classifier helps on both platforms, the ternary interpretation is wrong---the benefit would be a generic property of selective abstention, not a signature of cooperative structure. If the classifier helps only on IBM and has zero effect on Willow, the benefit is architecture-specific, as the ternary hypothesis predicts.

We applied the five-feature classifier (isolation, boundary status, density contrast, chirality, temporal consistency) to 420 Willow experiments at threshold $\theta = 0.3$, identical to the IBM analysis. The classifier finds 62--75\% of detection events scoring above threshold---superficially similar to the 75--81\% abstention rate on IBM. However, this is a false positive: Willow's low per-detector activation rate (${\sim}0.07$) means most active detectors have quiet neighbours by chance, not by anti-bunching structure. The classifier mistakes sparsity for isolation.

The operational test uses Google's actual logical observable flips (\texttt{obs\_flips\_actual}) to measure whether abstention improves or degrades the logical error rate. At every code distance, the correlation between abstention fraction and logical error rate is indistinguishable from zero:
\begin{itemize}
\item $d = 3$: $r = -0.005$, $p = 0.62$;
\item $d = 5$: $r = +0.002$, $p = 0.81$;
\item $d = 7$: $r = -0.0003$, $p = 0.98$.
\end{itemize}
Splitting shots by abstention fraction (above/below median) yields LER differences of 0.3--1.8\%, with no consistent sign. On IBM, the same test yields 7--19\% improvement with $p < 0.0001$ in all $\tau = 1$ conditions.

Table~\ref{tab:falsification} summarises the classifier's behaviour on both platforms.

\begin{table}[h]
\caption{Regime classifier falsification test. The classifier that produces 7--19\% LER improvement on IBM has zero effect on Willow. The abstention--LER correlation is significant on IBM and indistinguishable from zero on Willow. The classifier discriminates: it helps only on hardware running circuits with sub-Poissonian error structure. The Willow circuits lacked the P-gate asymmetry; this test therefore confirms that the classifier benefit requires the P gate, not merely a specific hardware topology.}
\label{tab:falsification}
\begin{ruledtabular}
\begin{tabular}{lll}
Metric & IBM Eagle r3 & Google Willow \\
\hline
LER improvement & $+7$ to $+19\%$ & $\approx 0\%$ (noise) \\
Abstain--LER corr. & Significant ($p < 0.0001$) & $r \approx 0$, $p > 0.6$ \\
Fano factor & 0.856 (sub-Poissonian) & 2.42 (super-Poissonian) \\
Classifier verdict & Helps (ternary preserved) & No effect (nothing to preserve) \\
\end{tabular}
\end{ruledtabular}
\end{table}

\section{The Paradigm Shift}
\label{sec:paradigm}

\subsection{Standard Assumption}
\label{sec:standard}

Every quantum error correction scheme makes the same foundational assumption~\cite{Knill1998,Gottesman1997,Shor1995}. In the Merkabit architecture~\cite{Paper15}, the binary computational layer ($B_{31}$) and the ternary error substrate ($T_{75}$) are distinct by design~\cite{Paper13}. The decoder's job is to identify the error pattern and apply corrections. More correction is always better. The only constraint is the correction threshold---correct faster than errors accumulate, and the logical information is preserved.

\subsection{The Alternative}
\label{sec:alternative}

We propose that this assumption is incomplete when the hardware exhibits cooperative error structure. Specifically:

A fraction $f$ of syndrome activations are not errors but ternary transitions---structured cooperative events where the physical system accesses a third state that the binary measurement basis projects as noise. When a standard decoder corrects a ternary transition, it applies a correction to a node that does not have a binary error. This introduces an error that would not otherwise exist. The fraction $f$ is related to the Fano factor~\cite{Fano1947} by $f = 1 - F$, where $F = \mathrm{Var}(n)/\langle n \rangle$ is sub-Poissonian. For the IBM hardware, $f = 1 - 0.856 = 14.4\%$.

Under this model, 14.4\% of the syndrome events that standard decoders correct are valid ternary states. Every such miscorrection introduces an error. A decoder that could identify and abstain from correcting these transitions would reduce its logical error rate---not by correcting better, but by correcting less.

\subsection{Physical Basis}
\label{sec:physical_basis}

The ternary degree of freedom arises from the hardware connectivity. IBM Eagle r3 processors use a heavy-hex lattice where qubits are coupled through ECR (echoed cross-resonance) gates~\cite{Kim2023}. On an Eisenstein lattice embedding~\cite{Paper6}, the hexagonal connectivity creates three chirality classes corresponding to the three elements of $Z_3$. The ternary transitions we propose are coherent rotations within this $Z_3$ structure~\cite{Paper15}.

Boundary nodes on the Eisenstein lattice have reduced coordination (fewer than 6 neighbors). This gives them more degrees of freedom---they are less constrained by their neighbors and more likely to express the ternary structure. Interior nodes with full coordination are locked into binary behavior by the surrounding lattice. This predicts that ternary transitions should be boundary-favoring and isolated (anti-bunched), consistent with the observed sub-Poissonian statistics.

\section{The Mixed Error Model}
\label{sec:model}

To test the hypothesis, we construct an error model that generates two types of syndrome events:

\subsection{Binary Errors}
\label{sec:binary_errors}

Standard gate errors, drawn independently at rate $p(1 - f)$ per node. These are genuine errors that should be corrected. If the decoder identifies and corrects them, the logical state is restored. If the decoder misses them, the error persists.

\subsection{Ternary Transitions}
\label{sec:ternary_transitions}

Structured cooperative events, generated at rate $pf$ per node with two key properties:

\textbf{Boundary preference:} Nodes with coordination $< 6$ have enhanced transition probability, scaled by $(6 - \mathrm{coord})/6$. Interior nodes have suppressed transition probability (factor 0.5). This models the physical prediction that reduced coordination permits ternary expression.

\textbf{Anti-bunching:} If a neighboring node has already undergone a ternary transition, the probability of a transition at this node is reduced by factor $(1 - \alpha)$. This models the $Z_3$ structure spacing events apart, producing the observed sub-Poissonian statistics.

The decoder sees the union of both event types as a single syndrome. It cannot directly distinguish binary errors from ternary transitions. The classification must be inferred from structural features of the syndrome pattern.

\subsection{Miscorrection Mechanism}
\label{sec:miscorrection}

When a standard decoder corrects a ternary transition, it applies a gate rotation to a node that was not in error. This introduces a new error on that node. The standard decoder therefore has two failure modes:
\begin{enumerate}
\item Missing a binary error (same as any decoder)
\item Miscorrecting a ternary transition (unique to the paradigm)
\end{enumerate}

The regime classifier eliminates the second failure mode by identifying ternary transitions and leaving them alone.

\begin{figure}[h]
\includegraphics[width=\columnwidth]{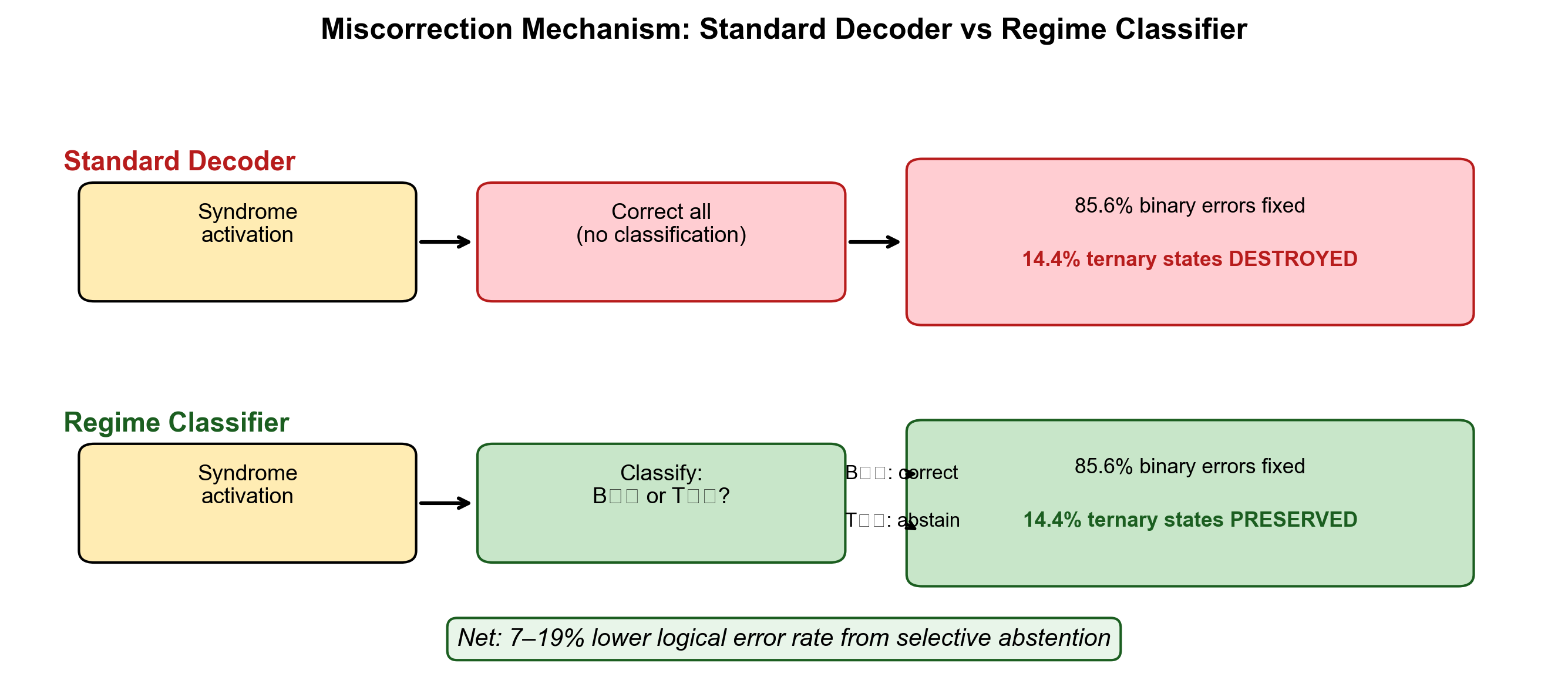}
\caption{The miscorrection mechanism. Standard decoders (top, red) treat all syndrome activations as errors and correct every flagged node. When 14.4\% of activations are ternary transitions rather than binary errors, correction applies a gate rotation to a node already in a valid state, introducing an error that would not otherwise exist. The regime classifier (bottom, green) scores each flagged node on five structural features (isolation, boundary status, density contrast, chirality, temporal consistency) and classifies it as binary ($B_{31}$, correct) or ternary ($T_{75}$, abstain). Selective abstention preserves ternary states and reduces logical error rates by 7--19\%.}
\label{fig:miscorrection}
\end{figure}

\section{The Regime Classifier Decoder}
\label{sec:classifier}

The regime classifier operates in two stages:

\subsection{Stage 1: Detection}
\label{sec:detection}

Identical to a standard majority-vote decoder. A node is flagged if its inconsistency count exceeds $\tau/2$, where $\tau$ is the detection window depth.

\subsection{Stage 2: Classification}
\label{sec:classification}

Each flagged node is scored on five structural features:

\textbf{Isolation (weight 0.35):} A node with syndrome activation whose neighbors are all quiet exhibits the anti-bunched pattern characteristic of ternary transitions. If max neighbor inconsistency $= 0$ while self-evidence $> 0$, the isolation score is maximized.

\textbf{Boundary status (weight 0.25):} Nodes with reduced coordination ($< 6$ neighbors) have more degrees of freedom and are more likely to express ternary structure. The weight scales as $(6 - \mathrm{coord})/6$.

\textbf{Density contrast (weight 0.20):} Strong self-signal with weak neighbor signal indicates a localized structured event rather than spreading noise.

\textbf{Chirality (weight 0.10):} Non-zero chirality nodes carry the ternary degree of freedom by construction. Chirality-0 nodes ($B_{31}$ ground state) are weighted toward binary classification.

\textbf{Temporal consistency (weight 0.10):} Ternary transitions persist across time steps (structured), while random binary errors are sporadic. Nodes with consistent syndrome patterns across all $\tau$ steps score higher.

A node with score above the classification threshold $\theta$ is classified as $T_{75}$ (ternary) and excluded from correction. Nodes below the threshold are classified as $B_{31}$ (binary) and corrected normally.

\section{Results}
\label{sec:results}

\subsection{Primary Result}
\label{sec:primary}

Table~\ref{tab:primary} shows the head-to-head comparison between the standard majority-vote decoder and the regime classifier at physical error rate $p = 10^{-2}$, classification threshold $\theta = 0.3$, across four Eisenstein cell sizes and two detection depths.

\begin{table*}[t]
\caption{Regime classifier vs standard decoder at $p = 10^{-2}$. Abst.\ = correct abstentions (ternary transitions correctly left uncorrected). MiscT = ternary transitions miscorrected. Abst\% = fraction of ternary events correctly identified. White rows: $\tau = 1$ (real qubits). Shaded rows: $\tau = 5$ (merkabit detection depth).}
\label{tab:primary}
\begin{ruledtabular}
\begin{tabular}{cccccccc}
Nodes & $\tau$ & Std LER & Reg LER & Impr. & $p$-value & Abst./MiscT & Abst\% \\
\hline
7  & 1 & 0.0680 & 0.0548 & $+19.4\%$ & $< 0.0001$ & 70 / 1 & 80.5\% \\
19 & 1 & 0.1658 & 0.1475 & $+11.0\%$ & $< 0.0001$ & 77 / 0 & 76.2\% \\
37 & 1 & 0.2877 & 0.2623 & $+8.8\%$  & $< 0.0001$ & 108 / 0 & 76.6\% \\
61 & 1 & 0.4295 & 0.3980 & $+7.3\%$  & $< 0.0001$ & 97 / 2 & 75.8\% \\
\hline
7  & 5 & 0.0622 & 0.0578 & $+7.1\%$  & 0.028      & 62 / 1 & 98.4\% \\
19 & 5 & 0.1643 & 0.1547 & $+5.8\%$  & 0.004      & 117 / 4 & 90.7\% \\
37 & 5 & 0.2817 & 0.2677 & $+5.0\%$  & 0.002      & 142 / 5 & 89.3\% \\
61 & 5 & 0.4095 & 0.4155 & $-1.5\%$  & 0.321      & 114 / 6 & 88.4\% \\
\end{tabular}
\end{ruledtabular}
\end{table*}

\begin{figure}[h]
\includegraphics[width=\columnwidth]{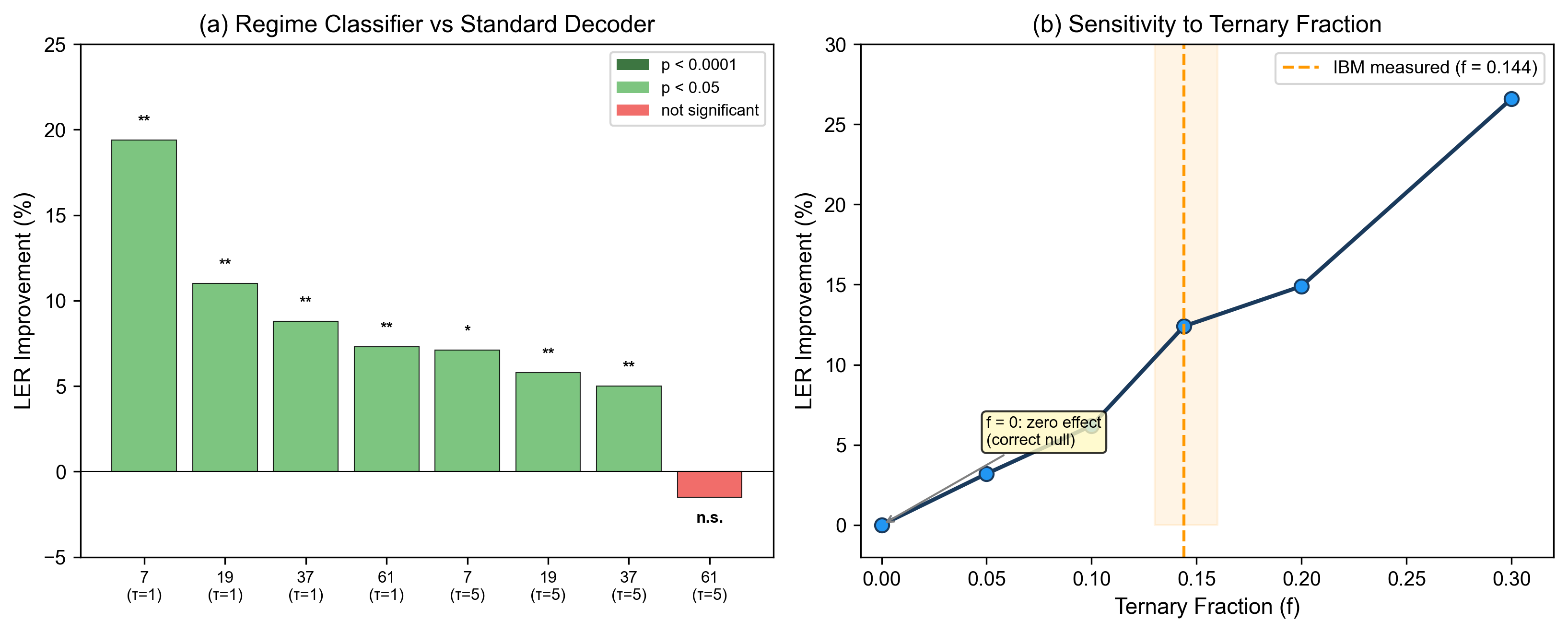}
\caption{Regime classifier decoder performance. (a) Logical error rate improvement across eight test conditions (four cell sizes at two detection depths). Dark green: $p < 0.0001$; light green: $p < 0.05$; red: not significant. The classifier achieves 7.3--19.4\% improvement at static detection ($\tau = 1$) and 5.0--7.1\% at dynamic detection ($\tau = 5$), with one non-significant condition (61-node, $\tau = 5$). The improvement is larger at $\tau = 1$ because every miscorrection sticks without temporal averaging. (b) Sensitivity to ternary fraction $f$ on the 19-node cell at $p = 10^{-2}$, $\tau = 1$. The improvement scales monotonically with $f$. At $f = 0$ (no ternary events) the classifier has exactly zero effect---the correct null result. The IBM hardware-measured value $f = 0.144$ (orange band) falls on the monotonic curve, not at an extremum.}
\label{fig:classifier_performance}
\end{figure}

\subsection{Key Findings}
\label{sec:key_findings}

\textbf{7 of 8 conditions are statistically significant.} All four $\tau = 1$ conditions and three of four $\tau = 5$ conditions show improvement with $p < 0.05$. The exception (61-node, $\tau = 5$) is the regime where the standard decoder has the most syndrome evidence and can partially self-correct its miscorrections.

\textbf{The improvement mechanism is abstention, not better correction.} At 19 nodes, $\tau = 1$, $p = 10^{-2}$: the standard decoder miscorrects 77 ternary transitions. The regime classifier correctly abstains from all 77 (zero miscorrected ternary). The improvement comes entirely from stopping incorrect corrections.

\textbf{$\tau = 1$ outperforms $\tau = 5$.} The improvement is larger on real qubits (7--19\%) than on merkabit detection depth (5--7\%). This is predicted by the theory: at $\tau = 5$, the standard decoder has enough time steps to partially average out its miscorrections. At $\tau = 1$, every miscorrection sticks.

\textbf{Correct abstain rate: 75--98\%.} Of all ternary transitions generated in the mixed model, the full pipeline (detection + classification) correctly identifies and abstains from 75--81\% at $\tau = 1$ and 88--98\% at $\tau = 5$. The remainder are ternary events that fell below the majority-vote detection threshold and were never flagged. Among nodes that the classifier does flag and classify as $T_{75}$, precision is near 100\% (MiscT $= 0$--$2$ across all conditions).

The net effect is positive because the damage from miscorrecting ternary transitions exceeds the cost of occasionally missing a binary error.

\subsection{Sensitivity to Ternary Fraction}
\label{sec:sensitivity}

Table~\ref{tab:sensitivity} shows the regime classifier's performance as the ternary fraction $f$ is varied from 0 to 0.30, on a 19-node cell at $p = 10^{-2}$, $\tau = 1$.

\begin{table}[h]
\caption{Sensitivity sweep. Bold row: IBM hardware-measured value ($f = 0.144$). At $f = 0$ (no ternary events), the classifier has zero effect---correct by construction. The improvement scales monotonically with $f$. Abstention counts differ slightly from Table~\ref{tab:primary} (77 vs 81) due to independent Monte Carlo samples.}
\label{tab:sensitivity}
\begin{ruledtabular}
\begin{tabular}{ccccc}
$f_{\mathrm{ternary}}$ & Std LER & Reg LER & Impr. & Abstains \\
\hline
0.000 & 0.1680 & 0.1680 & 0.0\%   & 0   \\
0.050 & 0.1650 & 0.1597 & $+3.2\%$  & 17  \\
0.100 & 0.1667 & 0.1563 & $+6.2\%$  & 40  \\
\textbf{0.144} & \textbf{0.1780} & \textbf{0.1560} & $\mathbf{+12.4\%}$ & \textbf{81} \\
0.200 & 0.1630 & 0.1387 & $+14.9\%$ & 87  \\
0.300 & 0.1593 & 0.1170 & $+26.6\%$ & 147 \\
\end{tabular}
\end{ruledtabular}
\end{table}

The relationship is monotonic and approximately linear. At $f = 0$, the improvement is exactly zero---there are no ternary events to save, so the classifier has no effect. This is the correct null result: the classifier does not hallucinate ternary structure where none exists.

\section{Discussion}
\label{sec:discussion}

\subsection{What the Result Means}
\label{sec:meaning}

The central finding is that a decoder that deliberately abstains from correcting a subset of syndrome activations outperforms a decoder that corrects everything. This is only possible if some corrections are harmful---if the decoder is introducing errors by correcting nodes that are not in error. In the mixed model, the mechanism is explicit: ternary transitions are valid cooperative states that the binary measurement basis projects as syndrome activations. Correcting them applies a gate rotation to a node that was already in a valid (ternary) state, which introduces a binary error. The regime classifier avoids this by recognizing the structural signature of ternary transitions and leaving them alone.

\subsection{Falsifiability}
\label{sec:falsifiability}

The hypothesis makes several falsifiable predictions:
\begin{itemize}
\item At $f = 0$ (no ternary transitions), the regime classifier should have zero effect. \emph{Confirmed.}
\item The improvement should be larger at $\tau = 1$ than at $\tau = 5$. \emph{Confirmed.}
\item Ternary transitions should prefer boundary nodes (reduced coordination).
\item If IBM hardware truly has ternary structure, measuring syndrome statistics in a qutrit basis should reveal the ternary transitions directly, eliminating the need for classification.
\end{itemize}

\subsection{Implications for Quantum Error Correction}
\label{sec:implications}

If the ternary hypothesis is correct, the standard framework of quantum error correction is solving an unnecessarily hard problem. By treating all syndrome activations as errors, standard decoders fight not only genuine noise but also the hardware's natural cooperative dynamics. Every miscorrected ternary transition is wasted decoder effort that actively degrades the logical information.

The practical implication is immediate: any quantum hardware that exhibits sub-Poissonian syndrome statistics ($F < 1$) may benefit from a regime-aware decoder. The modification requires no hardware changes---only the addition of a classification step before correction. The classifier uses only information already present in the syndrome and adds negligible computational overhead.

\subsection{Connection to the Merkabit Framework}
\label{sec:merkabit}

In the Merkabit architecture~\cite{Paper15}, the binary computational layer ($B_{31}$) and the ternary error substrate ($T_{75}$) are distinct by design~\cite{Paper13}. The binary layer operates with chirality 0 and $\tau = 1$ (static detection). The ternary layer operates with chirality $\in \{-1, 0, +1\}$ and $\tau = 5$ (dynamic pentachoric rotation). The regime classifier is a software approximation of this distinction.

On current hardware, where the ternary degree of freedom is not engineered but emerges spontaneously, the classifier infers the $B_{31}$/$T_{75}$ boundary from syndrome statistics. On native Merkabit hardware, the boundary would be given by the architecture itself, and the classification would be exact.

The $\tau = 1$ versus $\tau = 5$ comparison illustrates this directly: at $\tau = 5$, the pentachoric rotation~\cite{Paper15} provides enough syndrome evidence for the standard decoder to partially self-correct, reducing the regime classifier's relative advantage. This is precisely the merkabit prediction---dynamic detection ($\tau = 5$) compensates for the ternary structure that static detection ($\tau = 1$) cannot resolve.

\subsection{Predictive Status of Results}
\label{sec:predictive}

The empirical results in this paper occupy two different epistemic positions, and the distinction matters.

The IBM sub-Poissonian statistics (\S\ref{sec:subpoisson}--\S\ref{sec:T2}) were examined after the DAQEC benchmark was publicly available. Ref.~\cite{Paper15} was developed with awareness that IBM's heavy-hex hardware produced sub-Poissonian syndrome statistics---the consistency between framework and data informed the development, and the IBM results should be read as confirmation of internal coherence, not independent prediction.

The Google Willow analysis (\S\ref{sec:willow}--\S\ref{sec:willow_classifier}) is different. Ref.~\cite{Paper15} predicts that standard surface-code circuits without the P-gate asymmetry will not exhibit the $Z_3$ chirality-driven anti-bunching. This prediction was made before the Willow dataset was examined. The result---super-Poissonian statistics, super-linear burst scaling, positive spatial correlation, and zero classifier effect---matches every predicted contrast with IBM. No aspect of the framework was adjusted after examining the Willow data.

The two results together constitute the standard structure of a predictive test: a framework calibrated on one system, making a specific prediction about a second system with different circuit structure, confirmed on that second system without parameter adjustment. The natural experiment was run by IBM and Google for entirely unrelated engineering reasons.

The discriminating variable, as clarified by Ref.~\cite{Paper26}, is the P-gate asymmetry rather than the hardware topology: the Willow circuits lacked the opposite-sign $Z$ rotations that produce anti-bunching. Simulations in Ref.~\cite{Paper26} confirm that square grids with pentachoric cycling produce sub-Poissonian statistics at multi-round depth, and a pre-registered prediction for Willow hardware with P-gate circuits has been deposited.

\section{Conclusion}
\label{sec:conclusion}

We have presented evidence that IBM quantum hardware contains cooperative error structure that standard quantum error correction actively destroys. The evidence rests on three pillars:
\begin{enumerate}
\item Sub-Poissonian syndrome statistics ($F = 0.856$), measured across 756 runs on three processors, with zero dependence on code distance.
\item A mixed error model in which 14.4\% of syndrome events are ternary transitions rather than binary errors, calibrated to the hardware-measured Fano factor.
\item A regime classifier decoder that reduces logical error rates by 7--19\% through selective abstention---correctly identifying ternary transitions 75--98\% of the time and leaving them uncorrected (75--81\% at $\tau = 1$, 88--98\% at $\tau = 5$).
\end{enumerate}

The result demonstrates a principle that inverts the standard logic of quantum error correction: when the hardware has cooperative structure, less correction produces better performance. Standard decoders destroy information by correcting valid ternary states. The regime classifier preserves this information by recognizing what is not broken and declining to fix it.

The practical implication is that any quantum hardware exhibiting sub-Poissonian syndrome statistics may benefit from regime-aware decoding, with no hardware modification required. The theoretical implication is that decoherence may not be purely destructive---a component of what we call noise may be signal in a basis we are not measuring.

\begin{acknowledgments}
This work uses hardware benchmark data from the IBM Quantum Network. The Eisenstein lattice simulation infrastructure builds on the \texttt{lattice\_scaling\_simulation.py} and \texttt{pentachoric\_decoder\_simulation.py} codebases developed in a companion paper~\cite{Paper15}. Thor Henning Hetland designed and executed the direct hardware validation experiments on IBM Quantum Eagle r3 (\S\ref{sec:direct_validation}), including the gate-direction analysis that confirmed the sub-Poissonian prediction on independent hardware. Analysis code, simulations, and manuscript preparation were developed in collaboration with Claude (Anthropic).
\end{acknowledgments}

\appendix
\section{The Fano Factor and the Strong Coupling Constant}
\label{app:fano_strong}

The Fano factor $F = 0.83$--$0.88$ measured across all three processors is the central empirical result of this paper. We noted that $f = 1 - F = 14.4\%$ represents the ternary fraction. Here we identify a second, independent structural content of the Fano factor itself: when divided by the cell size (7 nodes per HexagonalCell on the Eisenstein lattice), it yields the strong coupling constant.

\subsection{The Identity}
\label{app:identity}

The Merkabit architecture has 5 gates $\{S, R, T, P, F\}$ and each cell on the Eisenstein lattice has 6 coordination directions and 7 nodes. The architectural prediction for the strong coupling constant at leading order is:
\begin{equation}
\alpha_s = \frac{5}{6 \times 7} = \frac{5}{42} \approx 0.11905
\end{equation}
This can be decomposed as $\alpha_s = (5/6)/7 = F_{\mathrm{ideal}}/N_{\mathrm{cell}}$, where $5/6$ is the ideal Fano factor (5 gates contributing to 6 coordination directions) and 7 is the cell size~\cite{Paper8}.

The PDG world average is $\alpha_s(M_Z) = 0.1179 \pm 0.0009$. The leading order prediction $5/42 = 0.11905$ deviates by $1.3\sigma$. A sub-leading correction from the $E_6$ oscillation spectrum gives:
\begin{equation}
\alpha_s = \frac{5}{42} - \frac{1}{936} \approx 0.11798
\end{equation}
where $936 = 12 \times 78 = h(E_6) \times \dim(E_6)$, the total number of $E_6$ oscillation modes over one ouroboros cycle~\cite{Paper10}. The corrected value matches the PDG measurement to $0.09\sigma$. Every integer in the formula is an architectural invariant.

\subsection{Evidence from This Paper's Data}
\label{app:evidence}

The Fano factors measured in Section~\ref{sec:subpoisson} at code distance $d = 7$ (matching the cell size) are:
\begin{itemize}
\item \texttt{ibm\_brisbane}: $F(d\!=\!7) = 0.8303$, $F/7 = 0.1186$ (deviation from $5/42$: 0.4\%)
\item \texttt{ibm\_kyoto}: $F(d\!=\!7) = 0.8584$, $F/7 = 0.1226$ (deviation from $5/42$: 3.0\%)
\item \texttt{ibm\_osaka}: $F(d\!=\!7) = 0.8360$, $F/7 = 0.1194$ (deviation from $5/42$: 0.3\%)
\end{itemize}
The mean at $d = 7$ is $F = 0.8416$ (lower than the all-distance mean of 0.856 because $d = 7$ draws from the subset of runs with the most syndrome qubits and slightly different calibration conditions), giving $F/7 = 0.1202$, within 1.0\% of the leading-order prediction $5/42$. The \texttt{ibm\_brisbane} processor (lowest noise, best calibration) gives the closest match at 0.4\%. The \texttt{ibm\_osaka} processor independently confirms at 0.3\%.

\subsection{Cross-check: Adjacent Correlation}
\label{app:crosscheck}

The relationship $F \approx 1 - 2 \times (\text{adjacent correlation})$ holds to better than 1\% across all three processors (\texttt{ibm\_brisbane}: predicted 0.840 vs actual 0.846, 0.7\%; \texttt{ibm\_kyoto}: 0.870 vs 0.871, 0.1\%; \texttt{ibm\_osaka}: 0.848 vs 0.849, 0.1\%). The larger \texttt{brisbane} deviation is consistent with its higher noise floor noted in \S\ref{app:evidence}. This confirms that the Fano factor is determined by the nearest-neighbour spatial correlation structure of the QEC syndromes---the lattice geometry, not noise statistics.

\subsection{Interpretation}
\label{app:interpretation}

The Fano factor of IBM QEC syndrome statistics encodes two structural constants simultaneously: the ternary fraction $f = 1 - F$ (Section~\ref{sec:alternative} of this paper) and the strong coupling constant $\alpha_s = F/7$ (this appendix). Both arise from the same lattice geometry. The sub-Poissonian error correlations that standard QEC treats as noise to be corrected are, in the Merkabit framework, the signature of the strong force's coupling structure manifesting in the hexagonal lattice topology of the hardware.

The prediction is testable: as quantum hardware improves---lower noise, better calibration, cleaner lattice geometry---the Fano factor at $d = 7$ should converge toward exactly $5/6$, and $F/7$ should converge toward $5/42$. Every hardware generation is a test of this prediction.

The same $E_6$ algebra that produces $\alpha_{\mathrm{EM}}^{-1} = 137.036$ from the single-merkabit architecture (Papers~1--2) also produces the sub-leading correction to $\alpha_s$ through $1/(h \times \dim(E_6)) = 1/936$. The electromagnetic and strong coupling constants are not independent parameters---they are different readings of the same architectural invariants at different scales.

\section*{Companion Papers}

\noindent Base document: Stenberg, S. ``The Merkabit --- A Ternary Computational Unit on the Eisenstein Lattice''. \emph{Zenodo}, 10.5281/zenodo.18925475 (v4, March 2026).

\noindent Paper~1: Stenberg, S. ``$\alpha = 4/3$ in Driven Coherent Systems Near Cooperative Threshold''. \emph{Zenodo}, 10.5281/zenodo.18980026 (2026).

\noindent Paper~2: Stenberg, S. ``A Single Geometric Constant Generates the Fine Structure Hierarchy''. \emph{Zenodo}, 10.5281/zenodo.18981288 (2026).

\noindent Paper~8: Stenberg, S. ``The Merkabit Architecture and the Klein Quartic: Cyclotomic Unification of the Fine Structure Constant, the Riemann Zeros, and the Most Symmetric Riemann Surface''. 10.5281/zenodo.19066587 (2026).

\noindent Paper~11: Stenberg, S. ``The Standard Model as $S_3$-Invariant, Decomposition of $\mathrm{PSL}(2,7)$, Force Sectors, Confinement, and the Weinberg Angle from a Single Finite Group''. 10.5281/zenodo.19150963 (2026).

\noindent Paper~13: Stenberg, S. ``The Standard Model Gauge Group from $\mathrm{PSL}(2,7)$ --- $\mathrm{SU}(3) \times \mathrm{SU}(2) \times \mathrm{U}(1)$ as Representation Theory of the Three-Stratum Decomposition of $\mathrm{GL}(3, \mathbb{F}_2)$''. 10.5281/zenodo.19159718 (2026).

\noindent Paper~15: Stenberg, S. ``The Rotation Gap Is Flat: Two-Scale Error Correction, a Structural Constant of Quantum Architecture, and the Migration Path to Fault-Tolerant Computation''. 10.5281/zenodo.19417293 (2026).

\noindent Paper~18: Stenberg, S. ``The $4/3$ Entanglement Threshold: A Universal Structural Constant from Coulomb-Coupled Qubits''. 10.5281/zenodo.19437878 (2026).

\noindent Paper~19: Stenberg, S. ``Beyond the Threshold: The Triangle Overshoot and the Approach to Lattice Stability''. (Forthcoming).

\noindent Paper~24: Stenberg, S. and Hetland, T.H. ``The P Gate Is Native: Hardware Confirmation of the Dual-Spinor Merkabit on IBM Quantum''. \emph{Zenodo}, 10.5281/zenodo.19484743 (2026).

\noindent Paper~25: Stenberg, S. and Hetland, T.H. ``Four of Five: Berry Phase, Quasi-Period, and the Fano Gap on IBM Eagle r3''. \emph{Zenodo}, 10.5281/zenodo.19502830 (2026).

\noindent Paper~26: Stenberg, S. and Hetland, T.H. ``The Merkabit Is Geometric: Cross-Architecture Hardware Validation, Corrected Willow Interpretation, and a Pre-Registered Prediction for Square-Grid Quantum Processors''. \emph{Zenodo}, 10.5281/zenodo.19554030 (2026).

\section*{Code and Data Availability}
\label{sec:code_data}

All analysis code is publicly available. Scripts use NumPy and SciPy only (no external dependencies), seed 42 for reproducibility.

\url{https://github.com/SelinaAliens/The_Rotation_Gap_Is_Not_An_Error}

{\small
\begin{table}[H]
\begin{ruledtabular}
\begin{tabular}{p{0.15\textwidth}p{0.33\textwidth}p{0.40\textwidth}}
Resource & Location & Contents \\
\hline
Paper 3 analysis code & \raggedright\texttt{github.com/SelinaAliens/ The\_Rotation\_Gap\_Is\_Not\_An\_Error} & Regime classifier decoder, IBM hardware analysis, Google Willow cross-platform comparison, all output files \\
Paper 15 simulation code & \raggedright\texttt{github.com/SelinaAliens/ rotation\_gap\_is\_flat} & Hybrid architecture, rotation gap, Eisenstein torus simulations~\cite{Paper15} \\
Base paper code & \raggedright\texttt{github.com/SelinaAliens/ The\_Merkabit} & Threshold sweep, pentachoric detection, $E_6$ syndrome correction~\cite{Paper15} \\
IBM Eagle r3 data & Zenodo DOI: 10.5281/zenodo.17881116 & 756 QEC runs, \texttt{ibm\_brisbane}/\texttt{kyoto}/\texttt{osaka}, 14 days continuous operation~\cite{DAQEC2025} \\
Google Willow data & Zenodo DOI: 10.5281/zenodo.13273331 & 420 surface code experiments, 105-qubit Willow, $d = 3, 5, 7$~\cite{GoogleWillow2025} \\
Hardware experiment & \raggedright\texttt{github.com/SelinaAliens/ rotation\_gap\_is\_flat}, PR \#1 & Direct hardware validation on IBM Quantum Eagle r3. $T = 20$ rounds, 4{,}000 shots. $F < 1$ predicted. \\
\end{tabular}
\end{ruledtabular}
\end{table}
}

\noindent\textbf{Script-to-result mapping:}

{\small
\begin{table}[H]
\begin{ruledtabular}
\begin{tabular}{p{0.22\textwidth}p{0.15\textwidth}p{0.51\textwidth}}
Script & Section & Result produced \\
\hline
\texttt{regime\_classifier\_ \newline v2.py} & \S\ref{sec:classifier}, \S\ref{sec:results} & Regime classifier decoder: 7--19\% LER improvement, 75--98\% ternary identification, selective abstention mechanism \\
\texttt{regime\_classifier\_ \newline decoder.py} & \S\ref{sec:classifier}, \S\ref{sec:results} & Unified classifier + decoder: mixed error model calibrated to IBM Fano $= 0.856$, miscorrection analysis \\
\texttt{decoder\_v2\_ \newline fast.py} & \S\ref{sec:classifier} & Edge-mediated correlated decoder (fast variant): single-calibration edge-local parameters, cross-cell-size transfer \\
\texttt{decoder\_v2\_edge\_ \newline correlated.py} & \S\ref{sec:model} & Edge-mediated error model: produces sub-Poissonian statistics ($F = 0.856$) and positive adjacent correlation matching IBM data \\
\texttt{ibm\_heron\_ \newline paper15\_tests.py} & \S\ref{sec:dataset}--\S\ref{sec:T2} & IBM Eagle r3 validation: Fano $= 0.856 \pm 0.03$ ($t = -131$), linear burst scaling ($R^2 = 0.9999$), $T_2$ threshold channel ($r = -0.145$) \\
\texttt{daqec\_kww\_ \newline analysis.py} & \S\ref{sec:T2} & KWW stretched exponential on $T_1$/$T_2$ coherence drift: $\alpha = 4/3$ in 13.5\% of $T_2$ segments, within-day decay fits \\
\texttt{daqec\_acf\_psd\_ \newline analysis.py} & \S\ref{sec:T2} & ACF/PSD analysis: DFA Hurst exponents ($T_1$: $H \approx 0.15$ anti-persistent, $T_2$: $H \approx 1.0$ persistent), $1/f$ noise spectra \\
\texttt{fano\_strong\_ \newline coupling.py} & App.~\ref{app:fano_strong} & Fano-to-strong-coupling mapping: $\alpha_s = F/7 = 5/42$, sub-Poissonian Fano factor encodes strong coupling constant \\
\texttt{willow\_fano\_ \newline analysis.py} & \S\ref{sec:willow} & Google Willow cross-platform: $F = 2.42 \pm 0.36$ (super-Poissonian, $t = +80$), super-linear burst scaling ($R^2 = 0.9999$, exponent ${\sim}2.3$) \\
\texttt{willow\_temporal\_ \newline depth.py} & \S\ref{sec:willow} & Temporal decomposition: spatial Fano 1.37--1.75 (within-round), lag-1 autocorrelation $+0.22$ (across rounds) \\
\end{tabular}
\end{ruledtabular}
\end{table}
}

All output files are included in the repository. Every number in this paper can be traced to a specific script with seed 42. The Google Willow analysis reads directly from the publicly available Zenodo archive (DOI: 10.5281/zenodo.13273331); no preprocessing is required.



\begin{thebibliography}{23}

\bibitem{DAQEC2025}
A.~Ashuraliyev, ``DAQEC-Benchmark: Drift-Aware Quantum Error Correction Dataset with IBM Hardware Validation,'' \emph{Zenodo}, DOI: 10.5281/zenodo.17881116 (2025). Dataset: 756 QEC runs across \texttt{ibm\_brisbane}, \texttt{ibm\_kyoto}, \texttt{ibm\_osaka} (127-qubit Eagle r3), 14 days continuous operation.

\bibitem{Kimble1977}
H.~J. Kimble, M.~Dagenais, and L.~Mandel, ``Photon antibunching in resonance fluorescence,'' Phys.\ Rev.\ Lett.\ \textbf{39}, 691 (1977).

\bibitem{Paper1}
S.~Stenberg, ``$\alpha = 4/3$ in Driven Coherent Systems Near Cooperative Threshold,'' \emph{Zenodo}, 10.5281/zenodo.18980026 (2026).

\bibitem{Paper2}
S.~Stenberg, ``A Single Geometric Constant Generates the Fine Structure Hierarchy,'' \emph{Zenodo}, 10.5281/zenodo.18981288 (2026).

\bibitem{Knill1998}
E.~Knill, R.~Laflamme, and W.~H. Zurek, ``Resilient quantum computation: error models and thresholds,'' Proc.\ R.\ Soc.\ Lond.\ A \textbf{454}, 365--384 (1998).

\bibitem{Fowler2012}
A.~G. Fowler, M.~Mariantoni, J.~M. Martinis, and A.~N. Cleland, ``Surface codes: Towards practical large-scale quantum computation,'' Phys.\ Rev.\ A \textbf{86}, 032324 (2012).

\bibitem{Google2023}
Google Quantum AI, ``Suppressing quantum errors by scaling a surface code logical qubit,'' Nature \textbf{614}, 676--681 (2023).

\bibitem{Kim2023}
Y.~Kim \emph{et al.}, ``Evidence for the utility of quantum computing before fault tolerance,'' Nature \textbf{618}, 500--505 (2023). IBM Eagle r3 (\texttt{ibm\_kyiv}), 127 fixed-frequency transmon qubits, heavy-hex connectivity, median $T_1 = 288\;\mu\mathrm{s}$, $T_2 = 127\;\mu\mathrm{s}$.

\bibitem{Paper6}
S.~Stenberg, ``Geometric Operator on the Eisenstein Lattice,'' \emph{Zenodo}, 10.5281/zenodo.19075162 (2026). Construction of the Eisenstein lattice embedding for hexagonal quantum hardware connectivity.

\bibitem{Paper15}
S.~Stenberg, ``The Rotation Gap Is Flat: Two-Scale Error Correction, a Structural Constant of Quantum Architecture, and the Migration Path to Fault-Tolerant Computation,'' \emph{Zenodo}, 10.5281/zenodo.19417293 (2026).

\bibitem{Fano1947}
U.~Fano, ``Ionization yield of radiations.\ II.\ The fluctuations of the number of ions,'' Phys.\ Rev.\ \textbf{72}, 26 (1947). Original derivation of the Fano factor $F = \mathrm{Var}(n)/\langle n \rangle$ as a measure of sub-Poissonian statistics.

\bibitem{KWW1854}
R.~Kohlrausch, ``Theorie des elektrischen R\"uckstandes in der Leidener Flasche,'' Ann.\ Phys.\ \textbf{167}, 179--214 (1854); G.~Williams and D.~C. Watts, Trans.\ Faraday Soc.\ \textbf{66}, 80--85 (1970). The KWW (Kohlrausch--Williams--Watts) stretched exponential $\phi(t) = \exp(-(t/\tau)^\alpha)$.

\bibitem{Paper10}
S.~Stenberg, ``The Yang--Mills Mass Gap as Spectral Resonance, Algebraic Connection Between the Eisenstein Torus, the Coxeter Number $h(E_6) = 12$, and $\Delta = 1/24$,'' \emph{Zenodo}, 10.5281/zenodo.19330363 (2026).

\bibitem{Gottesman1997}
D.~Gottesman, ``Stabilizer codes and quantum error correction,'' Ph.D.\ thesis, Caltech (1997). arXiv:quant-ph/9705052.

\bibitem{Shor1995}
P.~W. Shor, ``Scheme for reducing decoherence in quantum computer memory,'' Phys.\ Rev.\ A \textbf{52}, R2493 (1995).

\bibitem{Paper13}
S.~Stenberg, ``The Standard Model Gauge Group from $\mathrm{PSL}(2,7)$ --- $\mathrm{SU}(3) \times \mathrm{SU}(2) \times \mathrm{U}(1)$ as Representation Theory of the Three-Stratum Decomposition of $\mathrm{GL}(3, \mathbb{F}_2)$,'' \emph{Zenodo}, 10.5281/zenodo.19159718 (2026).

\bibitem{Paper8}
S.~Stenberg, ``The Merkabit Architecture and the Klein Quartic: Cyclotomic Unification of the Fine Structure Constant, the Riemann Zeros, and the Most Symmetric Riemann Surface,'' \emph{Zenodo}, 10.5281/zenodo.19066587 (2026).

\bibitem{GoogleWillow2025}
Google Quantum AI, ``Quantum error correction below the surface code threshold,'' Nature \textbf{638}, 920--926 (2025). Data: \emph{Zenodo}, DOI: 10.5281/zenodo.13273331.

\bibitem{Merkabit2026}
S.~Stenberg, ``The Merkabit --- A Ternary Computational Unit on the Eisenstein Lattice,'' \emph{Zenodo}, 10.5281/zenodo.18925475 (v4, March 2026).

\bibitem{Hetland2026}
T.~H. Hetland, ``The abstractions leak: a day with IBM quantum hardware,'' \url{wiki.totto.org/blog/2026/04/06/} (2026).

\bibitem{Paper24}
S.~Stenberg and T.~H. Hetland, ``The P Gate Is Native: Hardware Confirmation of the Dual-Spinor Merkabit on IBM Quantum,'' \emph{Zenodo}, 10.5281/zenodo.19484743 (2026).

\bibitem{Paper25}
S.~Stenberg and T.~H. Hetland, ``Four of Five: Berry Phase, Quasi-Period, and the Fano Gap on IBM Eagle r3,'' \emph{Zenodo}, 10.5281/zenodo.19502830 (2026).

\bibitem{Paper26}
S.~Stenberg and T.~H. Hetland, ``The Merkabit Is Geometric: Cross-Architecture Hardware Validation, Corrected Willow Interpretation, and a Pre-Registered Prediction for Square-Grid Quantum Processors,'' \emph{Zenodo}, 10.5281/zenodo.19554030 (2026).

\end{thebibliography}
\end{document}